\documentclass[10pt]{article}
\usepackage[letterpaper]{geometry}
\usepackage{hicss}
\usepackage{times}
\usepackage[none]{hyphenat}
\usepackage{url}
\usepackage{latexsym}
\usepackage{minted}
\usepackage{indentfirst}
\usepackage{graphicx}
\graphicspath{{images/}}
\usepackage[
    style=apa,
  ]{biblatex}
\usepackage{amsmath}
\usepackage{enumitem}
\usepackage{amsthm}
\usepackage{tcolorbox}
\usepackage{hyperref}
\usepackage{enumerate}
\usepackage{cleveref}
\usepackage{comment}
\usepackage{booktabs}
\usepackage[normalem]{ulem} %
\usepackage{paralist}

\newtcolorbox{ctfflagbox}{
  colback=gray!10,
  colframe=black,
  fonttitle=\bfseries,
  boxrule=0.5pt,
  arc=4pt,
  left=2pt,
  right=2pt,
  top=4pt,
  bottom=4pt
}
\usepackage{xcolor}
\definecolor{darkpink}{RGB}{219, 48, 122} %
\definecolor{darkblue}{rgb}{0.3, 0.3, 0.6}

\tcbset{
  myboxstyle/.style={
    colback=black!5,
    colframe=black,
    colbacktitle=black!79,
    boxrule=0.4pt,
    arc=1mm,
    left=6pt,
    right=6pt,
    top=6pt,
    bottom=6pt,
    fonttitle=\bfseries,
    width=\columnwidth,
  }
}

\tcbset{
  results_style/.style={
    colback=black!5,       %
    colframe=blue!50!black,      %
    colbacktitle=darkblue,  %
    boxrule=0.4pt,
    arc=1mm,
    left=6pt,
    right=6pt,
    top=6pt,
    bottom=6pt,
    fonttitle=\bfseries,
    width=\columnwidth,
  }
}

\newcommand{\change}[2]{{#2}}

\title{Evidence of Cognitive Biases in 
Capture-the-Flag Cybersecurity Competitions}

\author{
  Carolina Carreira \\
  Carnegie Mellon University, IST\\University of Lisbon, and INESC-ID \\
  \underline{carolinacarreira@cmu.edu}
  \And
  Anu Aggarwal \\
  Carnegie Mellon University \\
  \underline{anua@andrew.cmu.edu}
  \And
  Alejandro Cuevas \\
  Carnegie Mellon University \\
  \underline{acuevasv@andrew.cmu.edu}
  \AND
  Maria José Ferreira \\
  Carnegie Mellon University \\
  \underline{mariajor@andrew.cmu.edu}
  \And
  Hanan Hibshi \\
  Carnegie Mellon University and\\King Abdulaziz University \\
  \underline{hhibshi@andrew.cmu.edu}
  \And
  Cleotilde Gonzalez \\
  Carnegie Mellon University \\
  \underline{coty@cmu.edu}
}

\date{}

\begin{document}
\maketitle
\begin{abstract}
Understanding how cognitive biases influence adversarial decision-making is essential for developing effective cyber defenses. Capture-the-Flag (CTF) competitions provide an ecologically valid testbed to study attacker behavior at scale, simulating real-world intrusion scenarios under pressure. We analyze over 500,000 submission logs from picoCTF, a large educational CTF platform, to identify behavioral signatures of cognitive biases with defensive implications. Focusing on availability bias and the sunk cost fallacy, we employ a mixed-methods approach combining qualitative coding, descriptive statistics, and generalized linear modeling. Our findings show that participants often submitted flags with correct content but incorrect formatting (availability bias), and persisted in attempting challenges despite repeated failures and declining success probabilities (sunk cost fallacy). These patterns reveal that biases naturally shape attacker behavior in adversarial contexts. Building on these insights, we outline a framework for bias-informed adaptive defenses that anticipate, rather than simply react to, adversarial actions.

\end{abstract}

\subsubsection*{Keywords: Capture-The-Flag Competitions, Cognitive Bias, Mixed Methods, Cybersecurity}

\section{Introduction}

Cybersecurity is a high-stakes, asymmetric domain where defenders must secure complex systems while attackers need only exploit a single weakness. Adversaries often succeed not only through technical skill but also by exploiting human error, systemic blind spots, and novel vulnerabilities. Addressing this challenge requires not only stronger technologies but also a deeper understanding of adversarial thinking and behavior \parencite{rajivan2018creative}.

Capture-the-Flag (CTF) competitions are widely used for cybersecurity training and research \parencite{votipka2021hacked, chapman_picoctf_2014, marta2018, kees2017}. These events simulate realistic intrusion scenarios under time pressure, giving participants hands-on experience with cryptography, reverse engineering, and binary exploitation. Crucially, they require participants to adopt an attacker's mindset, making CTFs valuable testbeds for studying decision-making in adversarial contexts.

Most prior CTF research has emphasized technical performance, such as how participants solve problems, collaborate under pressure, or benefit from scaffolding. Yet, to strengthen defense, it is equally vital to understand the cognitive processes that shape attacker behavior. Prior studies emphasized the role of creativity, incentives, and strategy, but systematic deviations from rationality remain unexplored \parencite{rajivan2018creative, gutzwiller2015human, gutzwiller_oh_2018, gutzwiller2019cyber}. Behavioral science has documented such biases extensively \parencite{tversky1974judgment, kahneman2011thinking}, and early evidence suggests that attackers may also be influenced by availability and sunk cost fallacy \parencite{aggarwal2024evidence, ferguson2018tularosa}. However, most findings come from artificial lab settings with limited ecological validity.

In this paper, we analyze over 500,000 submission logs from picoCTF, one of the largest educational CTF platforms \parencite{Johnson2024PicoCTF}. This dataset offers a rare opportunity to observe attacker decision-making at scale in naturalistic conditions. We show that CTF environments can reveal psychologically meaningful patterns, particularly cognitive biases under realistic constraints, and propose ways to embed experimental manipulations in future CTFs to advance cognitive cybersecurity research. Specifically, our \textbf{main contributions} are threefold. First, we provide evidence that \textit{availability bias} and \textit{sunk cost fallacy} influence naturalistic attacker decision-making in CTFs. Second, we demonstrate large-scale, ecologically valid evidence using over 500,000 real-world submissions. Third, we outline a framework for embedding cognitive experimentation into CTF platforms to enable controlled studies in adversarial environments.

We next review related work on cognitive biases and CTFs, then describe our dataset and methodology for identifying bias signatures, before presenting results and discussing implications. %

\section{Background and Related Work}

Cognitive biases, systematic deviations from rationality, shape human decision-making, especially under pressure, uncertainty, or cognitive load \parencite{kahneman1982judgment}. Well-documented examples include the \textbf{availability bias}, a preference for easily recalled or recently encountered options \parencite{tversky_availability_1973}, and the \textbf{sunk cost fallacy} where individuals persist in low-payoff actions due to prior investment \parencite{ARKES1985124}. Across psychology, economics, political science, and medicine, such biases are shown to influence judgments and behaviors in predictable ways (e.g. \cite{tversky_loss_1991, knetsch_endowment_1989, nitu_improvising_2021}).
 
In cybersecurity, attackers operate in high-stress environments that make them susceptible to similar shortcuts. Recent work provides initial evidence. For example, red-team simulations show that deception techniques can induce confusion, frustration, and degraded performance among attackers \textcite{ferguson2018tularosa}. Other cyber range studies reveal sunk cost and default-setting biases \parencite{aggarwal2024evidence, aggarwal_adnd_2025, hitaj_case_2025}, while transcript analyses of red-team exercises identify anchoring, framing, confirmation, and sunk cost effects \parencite{gutzwiller_oh_2018,gutzwiller2019cyber}. 

Despite these insights, existing studies share some limitations. They rely on small samples or post hoc analyses, reducing generalizability. Some of the findings lack reproducible metrics for identifying biases in cyber contexts, and some are based on artificial and constrained laboratory settings, limiting ecological validity. As a result, the field still lacks systematic, large-scale evidence of how cognitive biases shape attacker decision-making in naturalistic adversarial contexts.

\subsection{Cognitive Bias in CTF Competitions}

CTFs are widely recognized as natural laboratories for studying attacker decision-making, especially when rich behavioral data (e.g., timestamps, submission histories, success/failure sequences, keystrokes) are available \parencite{Johnson2024PicoCTF, Savin_2023}. Originating with DEFCON in the late 1990s \parencite{cowan_defcon_2003}, CTFs have since become central to cybersecurity education, training, and research. CTF challenges (e.g., cryptography and reverse engineering) simulate offensive and defensive tasks under time pressure, requiring participants to probe weaknesses, test strategies, and adapt, making them well-suited for studying adversarial cognition and the potential emergence of cognitive biases.

Existing research has begun exploring this possibility. For example, \cite{yang_human_2025} designed targeted exercises to study loss aversion and satisfaction of search, while \cite{hitaj_case_2025} developed website exploitation challenges to probe representativeness bias. However, both approaches required interrupting gameplay or embedding artificial manipulations, limiting ecological validity.  Other work has examined team-level cognition: \cite{johnson_investigating_2022} identified biases such as groupthink, conformity, and escalation of commitment through transcript analysis, but did not analyze behavioral challenge data directly.  
Some researchers argue that CTFs' competitive elements, such as leaderboards and team dynamics, may distort behavior and yield incomplete insights into attacker decision-making \parencite{ferguson-walter_world_2019}. Yet others defend their value as controlled yet realistic environments to study biases and inform cyber defense \parencite{yang_human_2025, hitaj_case_2025, johnson_investigating_2022}. What remains missing is systematic evidence of how cognitive biases naturally emerge in large-scale CTF activity without intervention.

Our research addresses the current gap by analyzing behavioral logs from the picoCTF platform's practice environment, picoGym. By focusing on naturally occurring decision patterns in over 500,000 submissions, we examine how biases such as availability and sunk cost emerge under realistic adversarial conditions. We argue that establishing this naturalistic baseline is essential before embedding controlled manipulations, ensuring that future experiments reflect the contexts in which attacker cognitive biases genuinely arise.

\section{Methods}

This section describes our approach to detecting cognitive biases in CTF environments. Our study combines an exploratory qualitative review with quantitative modeling to identify patterns consistent with availability bias \parencite{tversky_availability_1973} and the sunk cost fallacy~\parencite{ARKES1985124}.

\subsection{The picoCTF Dataset}

PicoCTF is an open-source educational platform designed to promote cybersecurity education through gamified challenges \parencite{Johnson2024PicoCTF}. PicoCTF hosts annual competitions and provides a practice environment, picoGym\footnote{\url{https://play.picoctf.org/practice}}, where users can attempt challenges and access instructional materials. 

Our dataset comes from picoGym's internal logs, which record user activity and challenge submissions. The data is proprietary; collected, de-identified, and provided by the picoCTF development team. Although not publicly available, it can be obtained upon request to the administrators, subject to data use agreements and ethical approval.

Data was collected passively through the normal usage of the picoGym platform. The environment is browser-based, and users interact with challenges remotely. Each time a participant attempts to solve a challenge by submitting a flag, the platform records metadata about the submission. The data analyzed in our research spans a continuous period from \textbf{September 25, 2020, to April 8, 2024}, and it includes multiple picoCTF annual competitions and off-season usage. All submissions were logged under typical operating conditions without experimental manipulation.

The dataset contains \textbf{525{,}771 individual submission records} from \textbf{26{,}716 unique participants}, and 378 challenges. These challenges span six core categories: Web Exploitation, Cryptography, Reverse Engineering, Forensics, Binary Exploitation, and General Skills. Each row in the dataset represents a \textbf{single submission attempt} for a challenge by a user. Thus, the \textit{unit of analysis} is the individual flag submission.

\subsubsection{Data Preprocessing.}

To prepare the dataset for analysis, we removed duplicate user records to avoid inflating the number of participants or misrepresenting user-level activity. 
We also implemented a data processing pipeline to integrate raw submission logs with relevant challenge-level metadata. In particular, we joined each submission record with its corresponding challenge information to include contextual features such as the challenge category (e.g., Cryptography, Web Exploitation). 

\subsubsection{Ethical Considerations.}

No personally identifiable information was collected in the dataset, and all user identifiers were de-identified by the picoCTF prior to analysis and handing off the dataset to our research team. The dataset is considered anonymous to the authors of this paper because no member of the research team can re-identify the data.
All data used in this study and the research conducted were reviewed and approved by the Institutional Review Board. 

\subsubsection{Demographics.}

\change{}{Participants had the option to provide demographic information to picoCTF. Among those who did, a significant portion identified as male (72.9\%) and predominantly White (27.4\%), with most being from the United States (27.2\%). A comprehensive overview of the participants' demographics is presented in Table~\ref{tab:demographics}.}

\begin{table}
 \caption{Demographics of picoCTF participants (September 25, 2020, to April 8, 2024).}
 \label{tab:demographics}
 \small
 \centering
 \begin{tabular}{lrr}
  \toprule
  {} & \textit{n} & \% \\
  \midrule
  \multicolumn{3}{l}{\textit{Country}} \\
  United States   & 7,270 & 27.2\% \\
  India & 3,484 & 13.0\% \\
  United Kingdom    & 693 & 2.6\% \\
  Canada & 679 & 2.6\% \\
  Indonesia & 573 & 2.1\% \\
  Australia & 529 & 2.0\% \\
  Vietnam & 510 & 1.9\% \\
  Kenya & 501 & 1.9\% \\
  Other & 9,869 & 36.9\% \\
  Unspecified & 2,608 & 9.8\% \\
  \midrule
  \multicolumn{3}{l}{\textit{Gender}} \\
  Male  & 19,486 & 72.9\% \\
  Female & 3,234 & 12.1\% \\
  Unspecified & %
  3,908 & 14.6\% \\
  \midrule
  \multicolumn{3}{l}{\textit{Race/Ethnicity}} \\
  White or Caucasian & 7,325 & 27.4\% \\
  Asian & 7,264 & 27.2\% \\
  Other & 5,045 & 18.9.4\% \\
  Unspecified & 7,082 & 26.5\% \\
  
  \midrule
  \multicolumn{3}{l}{\textit{Player Type}} \\
  University & 13,013 & 48.7\% \\
  High School & 2,343 & 8.7\% \\
  Teacher & 971 & 3.6\% \\
  Other & 7,783 & 29.1\% \\
  Unspecified & 2,606 & 9.7\% \\
  \midrule
  \textbf{Total} & 26,716 & 100\% \\
  \bottomrule
\end{tabular}

\end{table}

\subsubsection{Key Variables in the Analysis.}

The main fields in the dataset are summarized in Table~\ref{tab:dataset-schema}. These variables form the basis for our behavioral and performance analysis.

\begin{table*}[h]
\centering
\caption{Key Variables in the PicoCTF Dataset}
\label{tab:dataset-schema}
\begin{tabular}{lp{7.5cm}l}
\hline
\textbf{Field Name} & \textbf{Description} & \textbf{Example} \\
\hline
\texttt{id} & Flag submission id  & \texttt{31} \\

\texttt{user\_id} & An anonymized identifier for the user & \texttt{u\_8a1f7d} \\

\texttt{timestamp} & Date and time of the submission & \texttt{2023-03-18T14:42:00Z} \\

\texttt{challenge\_id} & Unique identifier for the challenge & \texttt{challenge\_105} \\

\texttt{category} & Challenge category & \texttt{Cryptography} \\

\texttt{is\_correct} & Submission correctness (1 = correct, 0 = incorrect) & \texttt{1} \\

\texttt{submission\_text} & Text of the flag  & \texttt{picoCTF\{...\}} \\

\texttt{US} & If the participant is from the US  & \texttt{True} \\

\texttt{gender} & Participant Gender  & \texttt{Female} \\

\texttt{race} & Race & \texttt{White} \\
\hline
\end{tabular}
\end{table*}

\subsection{Operationalization of Cognitive Biases Signatures}

Rather than imposing preconceived notions of cognitive bias on the picoCTF dataset, we first conducted an exploratory review of participant behavior and challenge materials. 
During this exploration, we identified trends consistent with the availability bias and the sunk cost fallacy. After identifying bias trends, we then developed targeted metrics to detect these biases and applied the metrics to the picoCTF dataset. 
In this subsection, we describe each of the biases, the methods we used to analyze them, and the metrics we developed. %

\subsubsection{Availability Bias.}

Availability bias refers to the tendency to evaluate the likelihood or correctness of an option based on how readily it comes to mind \parencite{tversky_availability_1973}. In the context of picoCTF, all flags follow a predefined format: \texttt{picoCTF\{\textit{FLAG}\}}. 
However, solving the challenge in some cases outputs a flag string that is not immediately submission-ready. 

Before submitting, participants must apply specific alterations, such as adding the \texttt{picoCTF\{\}} wrapper or adjusting character formatting. These additional steps can lead to potential mistakes.%
We can find these instructions in the challenge description or hints, which participants might overlook due to availability bias, leading to submission errors in those challenges. 
We used three analysis methods to identify the availability bias: (1) Qualitative Coding, (2) Descriptive Statistics, and (3) Quantitative Statistics.

\textbf{Qualitative Coding}. We performed expert qualitative coding over all available challenges to capture the presence of such formatting instructions systematically. We evaluated each challenge according to two binary criteria:
    (1) Wrapper Mention: Does the description or hint explicitly mention the need to use the \texttt{picoCTF\{\}} wrapper?
(2) Other Flag Transformation: Does the description or hint instruct participants to apply additional formatting to the flag (e.g., changing case, removing spaces, modifying characters)?

A single primary coder reviewed all 378 challenges and applied these two binary labels based on the textual content of both the description and hint fields. To ensure coverage and accuracy, a second reviewer independently reviewed the labeled dataset to verify that no relevant challenges were missed. After all the data was coded, both coders met and discussed disagreements to reach a consensus. \change{}{As we followed a consensus-based approach, and the coders were experts, there was no need to calculate Inter-Rater Reliability~\parencite{mcdonald2019reliability}.}

\begin{tcolorbox}[myboxstyle, title=Availability Qualitative Coding Metric]
We use a binary metric to capture whether a challenge included explicit, non-standard instructions about the flag format. This metric indicates potential bias as deviations from the default format add cognitive load and are unrelated to the core technical task. 
\end{tcolorbox}

\change{}{
We frame this as availability bias because participants disproportionately rely on the most frequently reinforced flag format (\texttt{picoCTF\{\}}), which dominates their mental model. When instructions diverge from this norm, participants overlook them, not necessarily due to inattention, but because the standard wrapper is more cognitively available.}

\textbf{Descriptive Statistics.} Our qualitative analysis identified a subset of challenges where the output of solving the problem is not directly submission-ready: participants must apply specific formatting steps before submitting the flag. 
However, this may extend beyond the challenges that include explicit formatting instructions. 
As mentioned before, all flags have to be submitted using the \texttt{picoCTF\{\}} wrapper. However, many participants would know the content of a correct flag but forget to include the required formatting, the \texttt{picoCTF\{\}} wrapper (even when this is not mentioned in the description of the problem). This would suggest that participants correctly identified the core solution (the ``flag content'') but reverted to an incorrect submission format due to mental shortcuts.

We used the correct submissions to identify the correct \textit{flag\_core} per challenge by extracting the contents of the \texttt{picoCTF\{\}} wrapper and removing any trailing \textit{hex} suffix. We did this using a regular expression that split on the final underscore and verified that the suffix matched the hexadecimal format (e.g., \textit{flag\_fcac75c0} into  \textit{flag}). 
We excluded all incorrect submissions that included a \texttt{picoCTF\{\}} wrapper. This filtering ensured that our analysis focused exclusively on formatting errors (i.e., submissions that lacked the required wrapper). Next, we compared unwrapped incorrect submissions to the extracted \textit{flag\_core} values for each challenge. If a submission followed the pattern \textit{flag\_core} + \textit{hex} with a valid hexadecimal suffix, we categorized it as semantically correct but syntactically invalid. When this happens, it indicates that the participant correctly found the flag core but submitted it without the proper wrapper. Finally, we aggregated these misformatted-but-correct submissions at the challenge level. For each challenge, we computed the proportion of misformatted-but-correct relative to the total number of incorrect submissions. Filtered out challenges that had dynamic flags that did not follow the format described above, and challenges with fewer than 50 incorrect submissions.  We then conducted descriptive analyses of the remaining 321 challenges that had formatting errors.%

\begin{tcolorbox}[myboxstyle, title=Availability Descriptive Statistics Metric]
We define our \textbf{bias metric} as the ratio of misformatted but correct flag submissions to the total number of incorrect submissions, where a misformatted-correct submission contains the correct flag content but lacks the required \texttt{picoCTF\{\}} wrapper format or any other reformatting required for flag submission. This metric directly measures the proportion of failures due to formatting issues rather than incorrect problem-solving.
\end{tcolorbox}

\textbf{Quantitative Analyses.} We also used the misformatted rate per participant to model a binomial‐family generalized linear model (GLM) with a logit link. We treated the number of misformatted, incorrect submissions per participant as successes out of the total incorrect submissions. The model included race, gender, and U.S. identification as predictors. We fit the model using binomial errors, and residual deviance suggested no overdispersion.

\change{}{
\textbf{Alternative metrics.} 
Behavioral work on availability bias typically operationalizes the bias via a survey style with a message that contains equally manipulated items/classes, and participants need to assess the frequency of one of the items/classes present in the message%
~\parencite{tversky_availability_1973, kahneman2011thinking}.
In cyber settings~\parencite{aggarwal_adnd_2025} proposes a similar assessment where the message focuses on vulnerabilities with high and low criticality.
We considered alternative operationalizations of availability bias. One option was to use the \emph{first-attempt misformatting rate}, which isolates errors that occur before any feedback. However, this measure is sparse (only one data point per user–challenge). Another possibility was to apply an \emph{edit-distance threshold} to count any near-miss submissions as evidence of availability. 
While broader, this may detect mistakes with partial decoding errors and requires arbitrary distance cutoffs. %
}

\change{}{\textbf{Validity.} Submitting a misformatted but semantically correct flag does not exclusively imply availability bias. Alternative explanations include inattentiveness or lack of familiarity with picoCTF's submission requirements. However, the systematic and repeated nature of such errors across participants and challenges suggests availability bias.
Submitting the right core content in the wrong format isolates formatting issues, while keeping the problems constant. Under time pressure, participants may submit the most accessible flag (the core string they just decoded) rather than the correct flag (core \emph{plus} wrapper/transform), a canonical availability bias pattern~\parencite{tversky_availability_1973, kahneman2011thinking}.
}

\subsubsection{Sunk Cost Fallacy.}

In addition to availability bias, we also found evidence of the sunk cost fallacy. This cognitive bias refers to the tendency to continue investing in a task or decision due to prior investments (e.g., time or effort), even when doing so no longer maximizes expected outcomes \parencite{ARKES1985124}. 
In the context of CTF challenges, this may manifest as participants persisting in solving a specific problem despite repeated failure and decreasing likelihood of success, driven more by their previous effort than by a rational assessment of payoff.
We used two analysis methods to identify the sunk cost fallacy: (1) Descriptive Statistics, and (2) Quantitative Statistics.

\textbf{Descriptive Statistics.} We computed the empirical probability of a correct submission, conditional on the number of prior failed attempts. For each flag submission per participant, we grouped attempts based on how many failures had occurred beforehand (restricted to 1–30 prior attempts). Then we calculated the proportion of submissions in each group that were successful.

\textbf{Quantitative Analyses.} We modeled the probability of a correct submission as a function of the number of prior incorrect attempts. Specifically, we fit a generalized linear mixed-effects model (logistic link), with random intercepts for both users and challenges to account for individual- and item-level variation. For each submission, we recorded the user ID, challenge ID, correctness of the submission, and the number of prior incorrect submissions for that user–challenge pair.

\begin{tcolorbox}[myboxstyle, title=Sunk Cost Metric]
We define the sunk cost metric as the change in the probability of a correct submission with each additional incorrect attempt, conditioned on prior failures. 
\end{tcolorbox}

\change{}{
\textbf{Alternative metrics.} 
Prior research on sunk cost fallacy and escalation of commitment has proposed several ways of operationalizing persistence. Classic experimental studies often measure the likelihood of continued investment after receiving negative feedback or diminishing returns in survey styles, often with biased and unbiased choices~\parencite{ARKES1985124}. In cybersecurity research, persistence has been captured by the frequency of repeated exploits on a system despite previous failures~\parencite{aggarwal2024evidence}. 
We considered alternative operationalizations of sunk cost behavior in our CTF dataset. One option was \emph{time-to-switch}, measuring how long participants continue on a challenge before moving to another. However, this requires start-time data that we do not have, as our data only has flag submissions. }

\change{}{
\textbf{Validity.}
Our metric captures sunk cost behavior by measuring persistence on a challenge despite repeated failures. However, not all repeated attempts indicate bias, participants may retry because they made progress or discovered new information. Nevertheless, similarly to availability bias, the consistent pattern of continued attempts under low success probability suggests sunk cost behavior.
}

\section{Results: Availability Bias}

\subsection{Qualitative Coding Results}

Out of the 378 challenges analyzed, 60 contained explicit instructions regarding how the flag should be formatted before submission. These instructions appeared either in the problem description, the hint field, or both. Among these 60 challenges, 57 explicitly mentioned the need to use the \texttt{picoCTF\{\}} wrapper. At the same time, 28 provided additional instructions for transforming the flagged content, such as converting characters, adjusting formatting, or decoding values. Some challenges contained both types of instructions.

We noted a range of transformation requirements for the 25 challenges that included other flag-related instructions. The most common instruction involved converting the decoded output into hexadecimal format.

\subsection{Descriptive Results}

Out of 378 unique challenges in the dataset, we detected 165 challenges (44\%) with at least one case of a misformatted-but-correct flag, that is, a flag submission that contains the correct core string but omits the required \texttt{picoCTF\{\}} wrapper or has wrong capitalization. While the mean rate of misformatted-but-correct submissions is relatively low (2\%), the median ratio is only 0.63\%, and three-quarters of challenges stay below about 1.8\%. However, some challenges show high proportions. \change{The top two challenges were ``Binary Exploitation'' challenges with 38\% and 27\% of misformatted-but-correct submissions.}{ The top two challenges were in the ``Binary Exploitation'' category with 38\% and 27\% of misformatted-but-correct submissions.} %

\begin{figure}[tbp]
    \centering
    \includegraphics[width=0.75\linewidth]{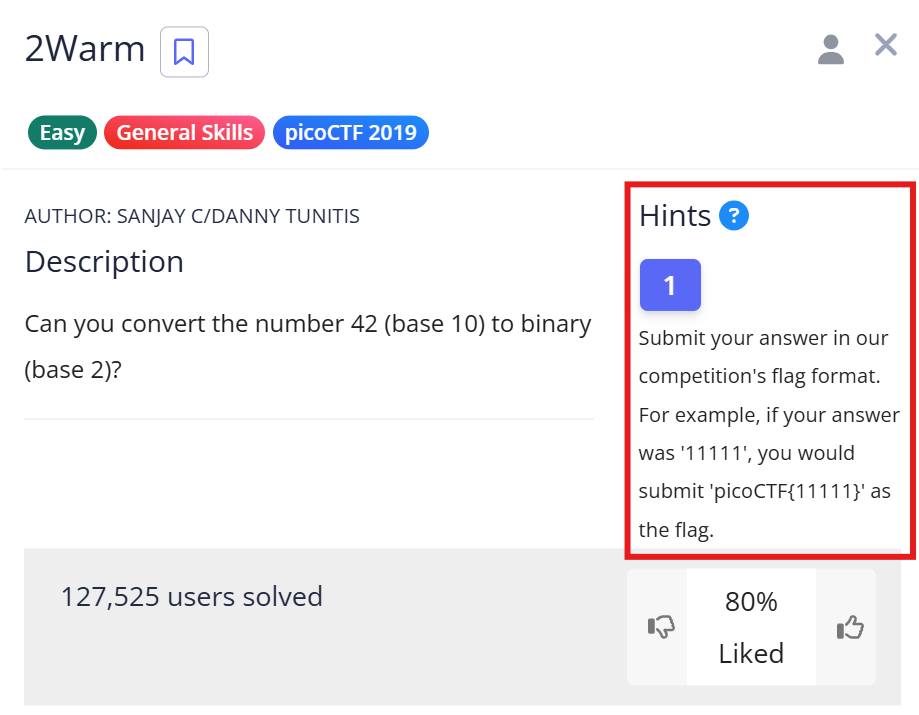}
    \caption{2Warm challenge from picoGym with hint showing the additional instructions for flag}
    \label{fig:2warm}
\end{figure}

To show an example of a challenge that has misformatted-but-correct flag submissions, we show the ``2Warm'' challenge. This challenge explicitly outlines the required format for flag submission in its description and hints, see Figure~\ref{fig:2warm}. The hints emphasize that the flag must be submitted in the format \texttt{picoCTF\{\}}. Upon successfully solving the challenge, participants receive the decoded flag, represented as ``101010''. Therefore, the appropriate format for submission is \texttt{picoCTF\{101010\}}.
Overall, 14\% of the ``2Warm'' incorrect submissions were misformatted-but-correct. 

\subsection{Quantitative Results}
We modeled misformatting propensity as a function of race, gender, and U.S. status using a binomial GLM. White (OR = 1.52, $p < .001$) and U.S. participants (OR = 1.52, $p < .001$) had significantly higher odds of submitting syntactically invalid flags compared to Asian and non-U.S. counterparts. Minority participants also showed elevated odds (OR = 1.21, $p < .001$), while Mixed-race participants did not differ significantly. Male participants had slightly higher odds than Non-male participants (OR = 1.08, $p = .020$). The baseline misformatting rate was $\approx2.9\%$, increasing to $\approx4.3\%$ for both White and U.S.-based users. No overdispersion was detected, supporting the binomial specification. Availability-bias errors seem to vary by demographic group.  Model fit was acceptable (AIC = 27{,}079).
These results suggest that availability bias in flag formatting is not uniformly distributed across demographics. Instead, specific user populations are more likely to commit errors indicative of this bias. The pattern may be due to different CTF conventions. We hypothesize that U.S.-based users, particularly White males, might be more likely to have prior experience with CTF competitions, and this may lead them to internalize default flag formats and deprioritize careful reading of instructions.

\begin{tcolorbox}[results_style, title=Availability Bias]
Our results suggest a recurring pattern in which participants recall or reproduce the correct flag core but neglect the required format, a signature of availability bias, where the most ``accessible'' version of the solution (e.g., partial string in memory or hints from others) is submitted, even when it does not satisfy the full format.
\end{tcolorbox}

\section{Results: Sunk Cost Fallacy}

\subsection{Descriptive Results}

\begin{figure}[htbp]
    \centering
    \includegraphics[width=0.9\linewidth]{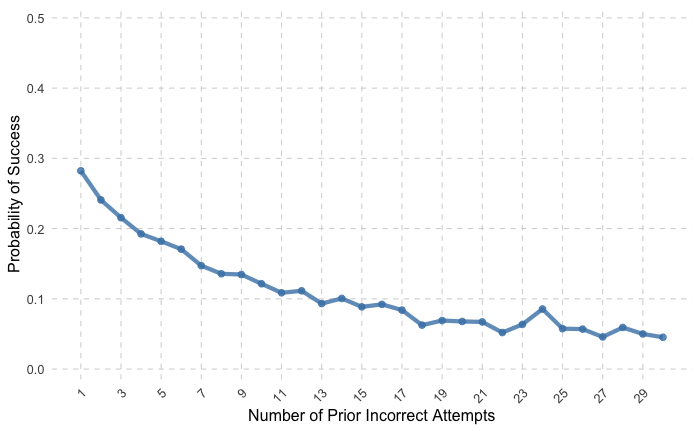}
    \caption{Empirical probability of a correct submission as a function of the number of prior incorrect attempts on the same challenge. Each point represents the average success rate observed among all submissions that occurred after a given number of failed attempts, aggregated across all users and challenges.}
    \label{fig:succ}
    \vspace{-5mm}
\end{figure}

Almost half of all submissions, 48.5\%, were successful on the first attempt (i.e., with zero prior failures). \Cref{fig:succ} displays the resulting success probabilities and reveals a strong negative association between the number of previous incorrect attempts and the probability of success.
This pattern is consistent with the presence of sunk cost behavior: users continue engaging with challenges even as the likelihood of success diminishes, possibly due to the psychological weight of prior investment. %

\subsection{Quantitative Results}

We ran a regression model to predict whether a challenge was eventually solved, using the number of incorrect attempts as a fixed effect and including random intercepts for both \texttt{user\_id} and \texttt{challenge\_id}. Our results showed a significant negative association between the number of incorrect attempts and the likelihood of solving a challenge ($\beta = -0.615$, SE = 0.003, $z = -205.23$, $p < .001$), so each additional incorrect attempt is associated with a substantial decrease in the odds of solving (OR $\approx 0.54$). The intercept ($\beta = 1.182$, SE = 0.052, $p < .001$) corresponded to an estimated 76.5\% probability of solving a challenge when no incorrect attempts have been made. Random effects indicated considerable variation in both user ability (SD = 0.495) and challenge difficulty (SD = 0.974), with greater variability between challenges than between users (AIC = 463{,}281.7; BIC = 463{,}326.3; log-likelihood = $-231{,}636.8$).

Despite this decline in success probability, users frequently continued to engage with the same challenge. The median number of incorrect attempts per user–challenge pair was three, and some users exceeded twenty. This pattern suggests persistence inconsistent with expected utility maximization and more aligned with a sunk-cost-driven motivation: users continue to invest effort into solving a challenge not because it becomes more solvable, but because they have already ``come this far''.

\begin{tcolorbox}[results_style, title=Sunk Cost Fallacy]
Our findings are consistent with the sunk cost fallacy as described in the behavioral decision-making literature. Participants appear to persist in their task engagement despite diminishing returns. %
In the CTF context, participants are free to switch tasks at any time, so this persistence may reflect non-normative decision strategies driven by prior effort rather than strategic optimization.\end{tcolorbox}

\section{Discussion}
\vspace{-1mm}

Cognitive biases shape human decision-making under uncertainty, with implications for both adversarial behavior and defensive system design. While prior research has examined attacker cognition, most studies rely on controlled simulations or hypothetical scenarios, limiting ecological validity \parencite{aggarwal2024evidence, aggarwal_adnd_2025, hitaj_case_2025}.

CTF competitions offer a promising alternative: they place participants in realistic, high-pressure problem-solving environments that simulate adversarial conditions, while generating detailed behavioral data. Yet, methods for identifying and validating cognitive biases in CTF environments remain underdeveloped. Our study takes a first step by demonstrating how availability bias and sunk cost fallacy can be observed in naturalistic picoCTF data and by proposing ways to leverage these insights for cyber defense.

\subsection{A Framework for Incorporating Biases into Defenses} 
\vspace{-2mm}

Inspired by the results of this study, we propose that future cybersecurity defense can leverage cognitive bias to detect attacker behavior and initiate adaptive defense mechanisms. Specifically, we envision systems that present carefully crafted stimuli (triggers) designed to exploit known cognitive biases. When an attacker responds in a way that indicates susceptibility, the system detects this via behavioral sensors and activates corresponding defenses.

Our findings suggest that attacker biases can be systematically operationalized for defensive purposes. We propose a three-part framework composed of bias triggers, sensors, and defensive mechanisms.
\begin{compactitem}
	\item 
    \textbf{Triggers} introduce stimuli designed to activate known cognitive bias.
	\item 
    \textbf{Sensors} detect behavioral signals that reveal susceptibility to bias.
	\item 
    \textbf{Defensive mechanisms} adapt system behavior to exploit these responses, divert attackers, or raise alerts. 
\end{compactitem}

CTF platforms can serve as testbeds for developing and evaluating this framework. By designing challenges with and without bias triggers, researchers can compare behavioral responses, estimate causal effects, and refine metrics for detecting biases. Combining in-game behaviors with standardized psychological instruments would further strengthen validation and generalizability. 

\subsection{Implications of Findings in the Cyber Domain}
\vspace{-2mm}

Our study illustrates how two well-known biases can inform defensive strategies (e.g., in future work):

\change{}{\textbf{Availability bias.} We observed frequent misformatted flag submissions, suggesting reliance on readily accessible but incorrect information (consistent with the definition of this bias \parencite{tversky_availability_1973}). In operational systems, defenders could deploy decoy files or directories with familiar names (trigger). Unauthorized access attempts (sensor) would then activate defensive mechanisms such as logging or redirection to monitored spaces. %
}

\textbf{Sunk Cost fallacy.} Persistent submissions following repeated failures reflect irrational continued investment. \change{}{Honeypots (i.e., strategically designed traps that present attackers with what seems to be critical information) can exploit this fallacy by presenting appealing but unproductive targets (trigger). Excessive repeated queries (sensor) could trigger defenses such as artificial delays or resource-draining tasks, consuming attacker time and energy.} 

These examples demonstrate how predictable cognitive patterns can be translated into actionable defenses, moving beyond reactive strategies toward adaptive, bias-informed security. Future work could translate these conceptual triggers, sensors, and defenses into deployable prototypes and test them in controlled CTF experiments.

\subsection{Limitations and Future Work}
\vspace{-2mm}

Our analysis is correlational. Observed patterns align with cognitive theory, but cannot confirm bias without experimental validation. Nonetheless, CTF environments offer a scalable and realistic foundation for developing tailored instruments to measure biases in cybersecurity contexts.

Data-related constraints also limit causal interpretation. PicoCTF logs provide rich submission records but lack fine-grained interaction data (e.g., clickstream data, in-browser hint use, or start times), preventing detailed modeling of effort or strategy. Also, metrics capture outcomes but not thought processes. \change{}{For example, the absence of sunk cost behavior may reflect true avoidance or an alternative optimization. Individual differences such as prior experience, domain knowledge, or motivation may further confound patterns. Additionally, although each player was assigned a unique participant ID, it remains theoretically possible for the same individual to register multiple accounts. However, this risk is mitigated by the fact that the picoCTF platform allows password recovery, reducing the incentive for duplicate registrations.}

\change{}{Another limitation concerns our metrics and data quality. While our measures capture behavioral outcomes, such as flag submissions, they provide no direct visibility into participants' motivations. For example, in the sunk cost fallacy, it is unclear whether players who did not show the bias were genuinely avoiding it or were instead pursuing a different strategic optimization. Similarly, because the data record only submission times and not the initiation time, we cannot directly link effort to elapsed engagement.} 

Future research should extend beyond availability and sunk cost to examine additional biases (e.g., default effect, anchoring, confirmation, and framing). Embedding controlled manipulations in ecologically valid CTFs, paired with surveys, reaction-time tasks, or adaptive interventions, would allow causal inference. Ultimately, advancing this integration of behavioral science and cybersecurity experimentation is essential for building adaptive defenses that anticipate, rather than merely react to, adversarial behavior.

\section{Conclusion}
\vspace{-2mm}

We investigated how cognitive biases influence decision‐making in CTF cybersecurity competitions by analyzing over 500,000 picoCTF submissions for signatures of availability bias and the sunk cost fallacy. We use a mixed-methods approach combining qualitative coding, descriptive statistics, and regression modeling. We found behavioral patterns consistent with these well-documented biases.

These findings contribute to cybersecurity research in three ways. First, they provide large-scale, ecologically valid evidence that attacker cognition is systematically shaped by biases, extending behavioral cybersecurity beyond small-scale lab studies. Second, they demonstrate the potential of CTF platforms as natural laboratories for studying adversarial decision-making, offering a scalable and realistic environment for behavioral analysis. Finally, they suggest new opportunities for bias-informed defenses: by embedding triggers and sensors into cyber systems, defenders may be able to detect and exploit predictable patterns in attacker behavior.

At the same time, our findings are limited by their correlational nature, providing only indirect evidence of cognitive processes. Looking forward, integrating behavioral science with cybersecurity experimentation can help build adaptive defenses that exploit predictable cognitive vulnerabilities as effectively as attackers exploit defenders' technical and cognitive weaknesses.

\section*{Acknowledgment} This research is based on work supported in part by the Office of the Director of National Intelligence (ODNI), Reimagining Security with Cyberpsychology-Informed Network Defenses (ReSCIND) program contract N66001-24-C-450, subcontract PO number: PO-0062554. This work was funded by the Portuguese Foundation for Science and Technology - FCT under PRT/BD/153739/2021 - CMU Portugal.

\printbibliography

\end{document}